\documentclass[twocolumn,showpacs,amsmath,amssymb,prl]{revtex4}

\usepackage{graphicx}
\usepackage{dcolumn}
\usepackage{bm}

\begin{document}


\title{Kondo effect in a few-electron quantum ring}

\author{U. F. Keyser}
\email{keyser@nano.uni-hannover.de}
\author{C. F\"uhner}
\author{S. Borck}
\author{R. J. Haug}
\affiliation{Institut f\"ur Festk\"orperphysik, Universit\"at Hannover, Appelstr. 2, 30167 Hannover, Germany}

\author{M. Bichler}
\author{G. Abstreiter}
\affiliation{Walter Schottky Institut, TU M\"unchen, 85748 Garching, Germany}

\author{W. Wegscheider}
\affiliation{Angewandte und Experimentelle Physik,  Universit\"at Regensburg, 93040 Regensburg, Germany}

\date{\today}

\begin{abstract}
A small quantum ring with less than 10 electrons was studied by
transport spectroscopy. For strong coupling to the leads a Kondo
effect is observed and used to characterize the spin structure of
the system in a wide range of magnetic fields. At small magnetic
fields Aharonov-Bohm oscillations influenced by Coulomb
interaction appear. They exhibit phase jumps by $\pi$ at the
Coulomb-blockade resonances. Inside Coulomb-blockade valleys the
Aharonov-Bohm oscillations can also be studied due to the finite
conductance caused by the Kondo effect. Astonishingly, the maxima
of the oscillations show linear shifts with increasing magnetic
field and gate voltage.
\end{abstract}


\maketitle

The characterization of semiconductor quantum dots by transport
spectroscopy is
an extremely successful approach
to understand the physics of interacting electrons confined to a
quasi-zero dimensional potential well~\cite{kouwenhoven-review}.
Until recently the only accessible shape of these quantum dots was
a tiny disc or box with a simple topology.  With new fabrication
techniques it is now possible to create more complex, multiple
connected topologies, namely small quantum rings with an outer and
an {\it inner} boundary.  These novel devices combine
characteristics of classical quantum dots with electronic
interference phenomena like the Aharonov-Bohm (AB)
effect~\cite{ab-effekt59}.

First small quantum rings were fabricated by self-assembled growth
of InAs on GaAs~\cite{lorke00}, but these structures were mainly
used for optical experiments.  An alternative approach, the local
oxidation of GaAs/AlGaAs-heterostructures with an atomic force
microscope (AFM)~\cite{ishii95}, allows to directly write tuneable
quantum rings into a two-dimensional electron gas
(2DEG)~\cite{fuhrer01,keyser02}.  These rings can be studied by
tunneling experiments.  Due to their small size, transport is
dominated by Coulomb blockade (CB)~\cite{kouwenhoven-review} and
the number of electrons on such a ring can be controlled by an
external gate voltage.  Recently, Fuhrer {\em et al.} studied a
quantum ring containing a few hundred electrons~\cite{fuhrer01}.
Their measurements showed an AB effect and allowed to deduce the
energy spectra of their device. Their system can be well described
within a single-particle picture~\cite{fuhrer01} because of an
effective screening of the electron-electron interaction by a
metallic top gate.

Here we discuss a small quantum ring containing less than ten
electrons in a totally different regime.  Due to the lack of a
screening top gate the ground state of our ring is dominated by
strong electron-electron interactions.  For a ring in this regime
a reduced AB period is predicted~\cite{niemela96}. We find indeed
such a reduction of the AB period in the transport measurements of
our small quantum ring. Furthermore, a strong coupling of the
AFM-fabricated device to the leads allows us to investigate a
Kondo effect~\cite{kondo64}.  We show that this well known
many-body phenomenon~\cite{glazman88,ng88,goldhaber98} is also
present in our system.  Our measurements show that electrons
confined on our quantum ring form a spin singlet state with
electrons in the leads.  The finite Kondo conductance allows us to
study the AB interference effects even in CB valleys. We report a
smooth shift of the AB oscillations with magnetic field and gate
voltage and compare it to recent results for a quantum dot
embedded into one arm of an AB ring~\cite{ji02}.

We fabricated our quantum ring from a $\delta$-doped
GaAs/AlGaAs-heterostructure containing a 2DEG 34~nm below the
surface. Details on the layer structure can be found in
Ref.~\cite{keyser02}. The 2DEG has an electron density of $n_e\sim
4\cdot10^{15}$~m$^{-2}$ and a mobility $\mu_e\sim
50$~m$^2$V$^{-1}$s$^{-1}$ at low temperatures. A detailed
description of our fabrication process was published
elsewhere~\cite{keyserAPL00,keyser02}.
\begin{figure}
\includegraphics{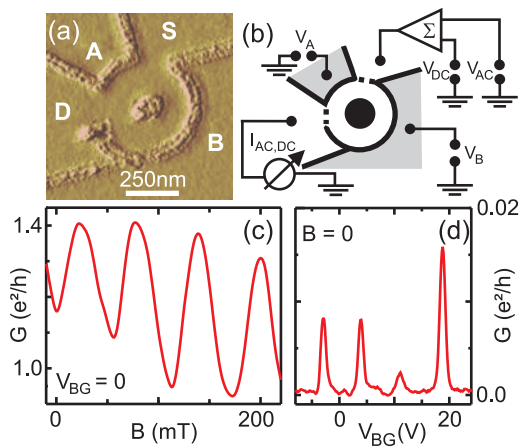}
\caption{(a) AFM image of the quantum ring with an inner diameter
of 190~nm and an outer of 450~nm. (b) Scheme of our measurement
setup. (c) Aharonov-Bohm oscillations in the open regime of the
quantum ring ($V_A,V_B>0$~mV). (d) Coulomb-blockade oscillations
as a function of the backgate voltage ($V_A=-250$~mV,
$V_B=-100$~mV, $B=0$).} \label{afmbild}
\end{figure}
An AFM-image of the completed ring structure is shown in
Fig.~\ref{afmbild}(a).  The two in-plane gates $A$ and $B$ are
separated from the ring by the rough oxide lines.  The ring is
connected to the leads by two 150~nm wide point contacts.  Both
point contacts are tuned by gate $A$ whereas gate $B$ couples only
to the source contact.  The inner diameter of the ring is 190~nm
and the outer diameter 450~nm.  In the following experiments, gate
$A$ is kept at a constant voltage $V_A$, and $V_B$ is used to
control the number of electrons on the ring.  Our measurement
setup is depicted in Fig.~\ref{afmbild}(b).  All measurements were
performed in a $^3$He/$^4$He-dilution refrigerator at a base
temperature $T_b\sim 30$~mK with an ac-excitation voltage of
5~$\mu$V at 89~Hz, added to a variable dc-voltage $V_{SD}$.  From
the temperature dependence of the CB peaks we deduce an effective
temperature of $\sim 50$~mK for the electronic system.

From the AB measurements shown in Fig.~\ref{afmbild}(c) we obtain
an AB period $\Delta B \sim 60$~mT at $V_A, V_B \geq 0$~mV for
electrons that are transmitted ballistically in a perpendicular
magnetic field.  The observed periodicity corresponds to a
diameter of $300$~nm for the electronic orbit, which fits
perfectly to the geometric values.  For measurements in the CB
regime we apply $V_A,V_B<-50$~mV to separate the ring from the
contacts by tunneling barriers. Typical CB oscillations as a
function of the voltage $V_{BG}$ applied to a metallic back gate
are shown in Fig.~\ref{afmbild}(d).  By analyzing CB diamonds from
nonlinear conductance measurements we extract a charging energy of
$U\sim 1.5$~meV and a single-particle level spacing of $\delta E
\sim 150$~$\mu$eV.

\begin{figure}[!t]
\includegraphics{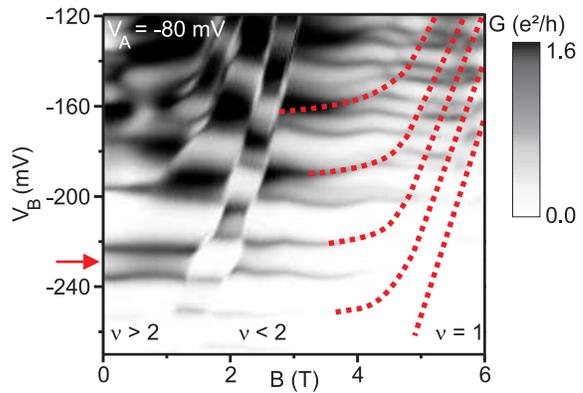}
\caption{Linear conductance $G$ through the ring in function of
the perpendicular magnetic field $B$ and the gate voltage $V_B$.}
\label{spectrum}
\end{figure}

The electron addition spectrum of our small quantum ring is shown
in Fig.~\ref{spectrum} for magnetic fields up to $B=6$~T.  The
linear conductance $G$ ($V_{SD}=0$) is plotted in grey scale as
function of $V_B$ and $B$ at $V_A=-80$~mV.  Each CB peak appears
as a black line more or less parallel to the $B$-axis.  Signatures
of non-vanishing conductance between CB peaks are observed and
attributed to the Kondo effect, e. g. as marked by the arrow in
Fig.~\ref{spectrum}.  If electrons are added according to Hund's
rule~\cite{tarucha96}, a Kondo effect is expected in consecutive
CB valleys~\cite{schmid00}.  In contrast, we observe an
alternation of the Kondo effect with electron number on the ring
(odd-even Kondo effect) below a gate voltage of -200~mV. For
$V_B>-200$~mV a more complicated pattern shows up at small $B$
presumably caused by the opening of the source tunneling barrier
and an increased asymmetry of the device.

An alternating pattern of high and low conductance is also
observed as a function of $B$.  The Kondo effect is modulated
abruptly between high (grey) and low (white) conductance regions
for $B < 2$~T.  Interestingly, our measurements
(Fig.~\ref{spectrum}) show some similarities with results obtained
for quantum dots designed as discs~\cite{schmid00,sprinzak02}.
This is presumably due to the similar importance of the outer edge
for tunneling through quantum dots as well as quantum rings in
high magnetic fields.  The alternating pattern of the valley
conductance with increasing magnetic field can be explained by a
redistribution of electrons between different Landau levels
(LL)~\cite{mceuen92}.  For example, inside the valley marked in
Fig.~\ref{spectrum} we observe an increased conductance indicating
a Kondo effect with an unpaired spin in the transport state at
small $B$.  At $B \sim 1.5$~T an electron from the upper LL is
transferred to the lower LL which is indicated by the sharp
boundary in the spectrum.  The Kondo effect is suppressed because
here the transport level in the lowest and outermost LL $n=0$
contains two electrons with opposite spins ($N=$odd, spin-0 in
$n=0$).  At $B\sim 2.0$~T a second electron is transferred to LL
$n=0$ and thus an unpaired spin is available again in LL $n=0$.
The Kondo effect is restored.  For higher $B$ similar drastic
changes are not observed anymore. We conclude that no further
electrons are redistributed from LL $n=1$ to $n=0$ and therefore
assume that all electrons are in the lowest Landau level (filling
factor $\nu=2$).

With a further increase of the magnetic field up to $B \sim 5.5$~T
we observe some weaker features highlighted in Fig.~\ref{spectrum}
by dashed lines.  These small variations in the amplitude and
position of the CB peaks are identified with spin flips of the
electrons in the lowest LL.  Their spin is flipped from
$|\uparrow>$ to $|\downarrow >$ by increasing $B$~\cite{ciorga00}.
For $B > 5.5$~T, the electrons in the lowest LL are totally
spin-polarized corresponding to a filling factor $\nu = 1$.  The
number of electrons $N$ on the ring is determined by counting the
spin flips between $\nu =2$ ($B=2$~T) and $\nu =1$ ($B\sim
5.5$~T).  For the marked Kondo valley, we observe only two spin
flips.  The occurrence of the Kondo effect at $B=0$ indicates that
$N$ is odd, thus we conclude that there are five electrons on the
quantum ring.  The same considerations apply for the surrounding
CB valleys, for which one extracts $N=4$ and $N=6$, respectively.
\begin{figure}
\includegraphics{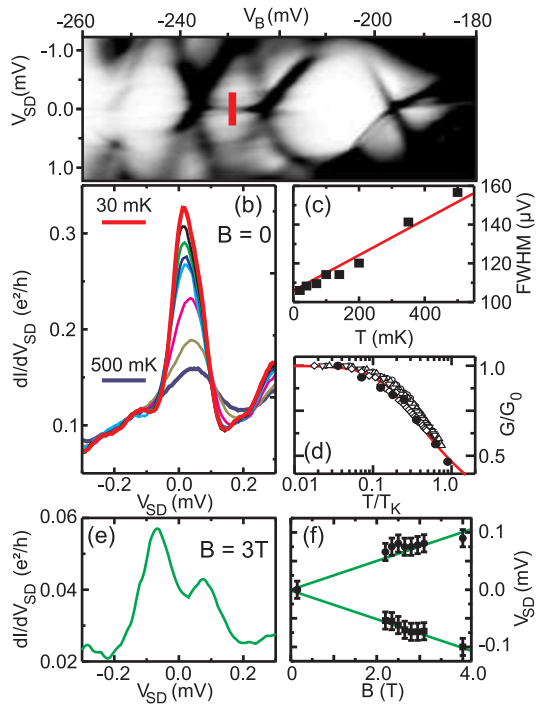}
\caption{(a) Grey scale plot of the differential conductance as
function of $V_B$ and source-drain voltage $V_{SD}$ at
$V_A=-80$~mV and $B=0$ with black (white) high (low) $G$.
(b)~Temperature dependence of the zero-bias peak in the center of
the $N=5$-diamond (marked by the short bar in (a)), $T$=30, 50,
70, 100, 140, 200, 350, 500~mK. (c) $T$-dependent full width at
half maximum (FWHM). (d) Peak conductance scaled with respective
$T_K$ and $G_0$ (see text) for data in (b) (full circles) and for
a stronger tunnel coupling (open symbols). (e) Split Kondo
resonance at $B=3$~T. (f) $B$-dependent positions of the split
Kondo peaks.} \label{nonlinearKondo}
\end{figure}

To analyze this Kondo effect in more detail, non-linear transport
measurements are shown in Fig.~\ref{nonlinearKondo}.
Fig.~\ref{nonlinearKondo}(a) depicts the differential conductance
$dI_{SD}/dV_{SD}$ measured at $T_b=30$~mK as function of $V_B$
and $V_{SD}$ ($V_A=-80$~mV, $B=0$).  The central CB diamond
corresponds to the valley discussed above.  A sharp zero-bias peak
appears whereas in the valleys to the left and right only low
conductance is observed. Fig.~\ref{nonlinearKondo}(b) depicts
temperature dependent measurements at the gate voltage marked in
Fig.~\ref{nonlinearKondo}(a).  The zero-bias peak observed at
$T_b=30$~mK vanishes almost completely when the temperature is
increased to $T\sim 500$~mK as expected for a Kondo
resonance~\cite{meir93}.  From these measurements we estimate a
Kondo temperature $T_K$ by extrapolation of the full peak width at
half maximum $\Delta V_{SD}$ to $T=0$~K
(Fig.~\ref{nonlinearKondo}(c)), which results in $T_K \approx {e
\Delta V_{SD,T=0}/2 k_B} \sim 600$~mK~\cite{meir93,nygard00}.
Fitting the Kondo conductance at zero bias $G(T)$ with an
empirical formula from
Ref.~\cite{costi94,goldhaberPRL98,nygard00}, $G (T) =
G_0/\left(1+(2^{1/s}-1)(T/T_K)^2\right)^s$ yields $T_K\sim 600$~mK
and $s\sim 0.21$.  This is in agreement with earlier studies on a
spin-1/2 system~\cite{goldhaberPRL98}.  The scaled peak
conductance (full circles) is shown together with the fit in
Fig.~\ref{nonlinearKondo}(d).  To prove the scaling behavior,
results obtained at four different gate voltages are shown as open
symbols in Fig.~\ref{nonlinearKondo}(d). The respective Kondo
temperatures are $T_K \sim 0.9, 1.0, 1.1, 1.4$~K.

The zero-bias peak of a spin-1/2 Kondo effect is expected to split
with increasing magnetic field according to the Zeeman effect,
$\Delta E = g_{GaAs} \mu_B B$ ($g_{GaAs}=0.44$, $\mu_B$ Bohr's
magneton). This peak splitting is shown at $B=3$~T in
Fig.~\ref{nonlinearKondo}(e).  The different peak amplitudes are
related to a slight asymmetry in the coupling of the ring to the
leads.  The peak positions in $V_{SD}$ extracted from several
measurements for 2~T~$<B<4$~T are plotted in
Fig.~\ref{nonlinearKondo}(f), the lines indicate the expected peak
positions.  We observe a nice agreement with the measurement for
$B>2$~T which is another evidence for a spin-1/2 Kondo effect.
Between 1.5~T and 2.1~T the Kondo effect is absent due to the
paired spin configuration.  For $B < 1.5$~T the peak splitting is
not resolved, presumably due to the broadening of the Kondo peak
of the order of $k_B T_K \sim 50$~$\mu$eV.

\begin{figure}
\includegraphics{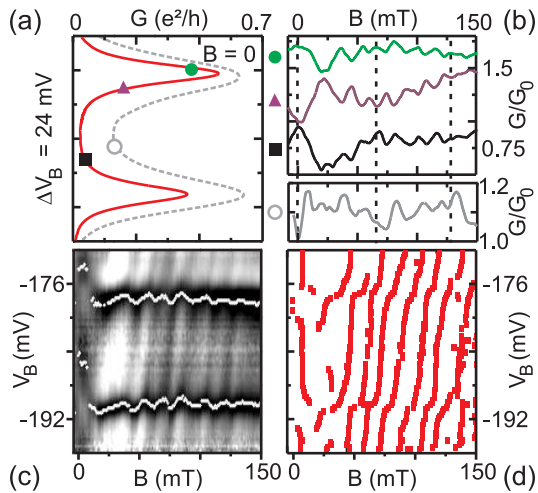}
\caption{(a) $N=5$-Kondo valley at $V_A=-150$~mV (solid) and
$V_A=-80$~mV (dashed line) at $B=0$. $\blacksquare,
\blacktriangle, \bullet$ and $\circ$ mark the gate voltages for
the Aharonov-Bohm measurements in (b). (b) Upper part: Normalized
conductance $G/G_0$ as a function of $B$ at $V_A=-150$~mV. The
curves are offset for clarity. Lower part: $G/G_0$ at $V_A=-80$~mV
after subtraction of an increasing background. (c) $G/G_0(B,V_B)$
as grey scale plot at $V_A=-150$~mV with peaks marked by white
dots. (d) Positions of the Aharonov-Bohm maxima extracted from
(c). } \label{Aharonov-BohmKondo}
\end{figure}

The CB peaks are broadened due to the strong coupling in the Kondo
regime as well (dashed line in Fig.~\ref{Aharonov-BohmKondo}(a),
$V_A=-80$~mV).  This broadening obscures the small shifts in the
peak positions as induced by changes of the ground states in the
ring.  To avoid this, the Kondo valley marked in
Fig.~\ref{spectrum} is shown again in
Fig.~\ref{Aharonov-BohmKondo}(a) at a slightly lower tunnel
coupling at $V_A=-150$~mV.   This reduces the conductance in the
Kondo regime to below $0.1 e^2/h$.  Due to the finite capacitance
between the ring and gate $A$, the center of the valley is shifted
from $V_B\sim -230$~mV to $V_B\sim -185$~mV.  In
Fig.~\ref{Aharonov-BohmKondo}(b) we show AB oscillations in the
normalized conductance $G/G_0$ as a function of $B$ for the gate
voltages marked by the symbols in
Fig.~\ref{Aharonov-BohmKondo}(a).  The vertical dashed lines
denote the expected period of $\Delta B \sim 60$~mT of the AB
oscillations extracted from the measurements in the open regime.
It is immediately evident that we obtain a much {\em shorter}
period.  This shorter period is in contrast to the results of
Fuhrer {\it et al.}~\cite{fuhrer01}, who obtained the {\em normal}
AB period for their ring with many electrons and screened
electron-electron interaction.  A comparable oscillation period is
also obtained in the center of the Kondo valley at higher tunnel
coupling $V_A=-80$~mV and $V_B=-231$~mV (marked by the open circle
in (a)) as depicted in the lower part of
Fig.~\ref{Aharonov-BohmKondo}(b).  The fast AB oscillations are
also reflected by the movement of the CB peaks in
Fig.~\ref{Aharonov-BohmKondo}(c) as well.  Each kink in the CB
peak position indicates a change of the ground state in our ring.

According to a theoretical calculation of Niemel\"a {\it et al.}
electron-electron interaction lifts the degeneracy of the singlet
and triplet states in a ring~\cite{niemela96}.  This effect leads
to more frequent level crossings of the ground state compared to a
ring without interaction because the triplet states are present at
lower energies.  These calculations predict a transition (AB)
period $\Delta B$ that shortens with increasing number of
electrons on the ring, e.g. for four electrons it should be four
times smaller.  For our ring with five electrons $\Delta B$ should
be shortened by a factor of five.  In fair agreement, we observe a
period of $\Delta B \sim 13$~mT which is four to five times
shorter compared to the open regime. The deviations from perfect
periodicity especially at small magnetic fields have to be
explained by the influence of some residual disorder.

We utilize the Kondo effect with its finite conductance to study
the evolution of the AB effect in the CB valley. In
Fig.~\ref{Aharonov-BohmKondo}(d) each square marks the position of
an AB maximum in the $(V_B,B)$-plane extracted from the data
presented in Fig.~\ref{Aharonov-BohmKondo}(c).  The small kinks,
visible for all magnetic fields at both CB peaks at
$V_B\sim-190$~mV and $V_B\sim -177$~mV indicate a phase jump of
$\pi$. This result is verified in
Fig.~\ref{Aharonov-BohmKondo}(b): the conductance at $B=0$ changes
from a maximum to a minimum ($\bullet$ to $\blacktriangle$) in
crossing the peak.
 Apparently, we observe smooth linear shifts of
the AB maxima in Fig.~\ref{Aharonov-BohmKondo}(d), indicated by
the slightly tilted vertical lines.  These appear in our
two-terminal measurement only at finite magnetic field when the
ring is threaded by at least half a flux quantum.  Linear shifts
of {\em normal} AB oscillations were recently reported by Ji~{\em
et al.} for a quantum dot in the unitary Kondo regime embedded in
an AB interferometer~\cite{ji02}.  Their four-terminal measurement
is interpreted in terms of smooth phase shifts by almost $2\pi$
across a Kondo resonance.  In contrast we investigate a Kondo
resonance far from the unitary limit in a quantum dot which itself
serves as the interferometer.  The exact mechanism for the
observed linear shift of the AB maxima is still to be clarified,
but it might be connected to the fact that the level structure of
our small ring interferometer is influenced by the gate voltage.
For a detailed understanding further theoretical work is
necessary.

In conclusion, a small tuneable quantum ring with less than ten
electrons is shown to exhibit Aharonov-Bohm oscillations as well
as Coulomb-blockade.  At strong coupling to the leads we find
evidence of a Kondo effect induced by a single spin on the ring.
The energy spectrum is strongly influenced by electron-electron
interaction.  An analysis of the phase evolution of the
Aharonov-Bohm effect in the Kondo regime yields phase jumps by
$\pi$ at the Coulomb-blockade resonances and a smooth shift of the
Aharonov-Bohm maxima in between.

We acknowledge discussions with S. Ulloa and J. K\"onig and
financial support by BMBF, DIP and TMR.


\end{document}